\documentclass[12pt]{article} 
\usepackage{epsfig}

\makeatletter 
\@addtoreset{equation}{section} 
\makeatother 
 
\newcommand{\beq}{\begin{equation}}
\newcommand{\eeq}{\end{equation}}
\newcommand{\be}{\begin{eqnarray}}
\newcommand{\ee}{\end{eqnarray}}

\renewcommand{\Im}{{\mathrm{Im}}}
\renewcommand{\Re}{ {\mathrm{Re}} }
\newcommand{\ov } {\over }
\newcommand{\p }{\partial }
\newcommand{\s }{\sigma }
\newcommand{\al }{\alpha }
\newcommand{\T }{ {\cal T } }

\def\appendix#1{
  \addtocounter{section}{1}
  \setcounter{equation}{0}
  \renewcommand{\thesection}{\Alph{section}}
  \section*{Appendix \thesection\protect\indent \parbox[t]{11.15cm}
  {#1} }
  \addcontentsline{toc}{section}{Appendix \thesection\ \ \ #1}
  }

\begin{document}

\begin{titlepage}

\begin{center}
\hfill hep-th/0210245\\
\hfill CERN-TH/2002-297\\
\hfill SISSA 72/2002/EP \\

\vskip  1 cm

\vskip 1 cm

{\Large \bf The decay of massive closed superstrings with maximum angular momentum}

\vskip  1 cm

{\large Roberto Iengo $^a$ and Jorge G. Russo$^{b,c}$}


\end{center}


\centerline{\it ${}^a$ International School for Advanced Studies (SISSA), Via Beirut 2-4,}
\centerline{\it I-34013 Trieste, Italy}
\centerline{\it  INFN, Sezione di Trieste }
\medskip

\centerline{\it {}$^b$ Theory Division, CERN, CH-1211 Geneva 23, Switzerland}

\medskip

\centerline {\it {}$^c$ Departamento de F\'\i sica, Universidad de
Buenos Aires and Conicet, }
\smallskip
\centerline {\it Ciudad Universitaria, Pab. I, 1428
Buenos Aires, Argentina}

\vskip 0.4 cm

\vskip 1.5 cm

\begin{abstract}

We study the decay of a very massive closed superstring (i.e. $\al' M^2\gg 1$)
in the unique state of maximum angular momentum.  
This is done in flat ten-dimensional spacetime and in the regime of weak string coupling,
where the  dominant decay channel is into two states of masses $M_1,\ M_2$.
We find that the lifetime surprisingly grows with the first power of the mass $M$:
$\T =c \al' \ M$.
We also compute the decay rate for each values of $M_1,\ M_2$.
We find that, for large $M$, the dynamics selects only special channels of decay:
modulo processes which are exponentially suppressed, 
for every decay into a state of given mass $M_1$, the mass $M_2$
of the other state is uniquely determined.


\end{abstract}

\bigskip
\bigskip
\bigskip

\date{October 2002}

\end{titlepage}

\newpage

\section{Introduction}

{}In type II superstring theory in flat, ten-dimensional non-compact spacetime,
all massive strings are generally expected to be unstable quantum mechanically by a decay into lighter particles. 
Massive strings are the key ingredient of string theory and crucial for the consistency of the theory.
Despite many years of study of string theory, very little is known about the  way a massive string decays.

Earlier calculations of decay properties 
are  in \cite{Green,Mitchell,Dai,Okada,Sundborg,Turok}, and more recent
studies can be found in \cite{Amati2,IK,Manes}.
A calculation for states with maximum angular momentum
in the open string theory was given in \cite{Turok}.
An inclusive decay rate was computed in \cite{Amati2}.
A more recent calculation for open and closed superstring
theory for {\it generic } states (which have angular momentum 
much less than the maximum value)
 is done in \cite{Manes}.

{} For small string coupling, the dominant elementary process is the decay
into two particles. If these particles are massive, then each of them will subsequently 
decay into two ligther particles, and the process ends
when only massless particles remain.

Here we shall explicitly compute the rate for  the decay of the massive string 
states of maximal angular momentum into 
two particles.  This will be done for the closed (type IIA or IIB)
superstring theory 
in flat ten-dimensional spacetime.
We will consider all cases: when these particles are both massive, 
when one of them is massless, and when both of them are massless.

A general formula for the decay can be obtained from a one-loop calculation
of the mass-shift.
The inverse lifetime $\T^{-1}$ of a massive state of mass $M$ is then given by 
\beq
\T^{-1}=\Gamma = {\Im \Delta M^2 \over 2M}~.
\label{aaq}
\eeq
where $\Delta M^2$ represent the radiative correction to the mass, to be
expressed as a loop expansion. At one loop, it receives a
contribution from the
two-particle intermediate states, and thus $\Gamma$ is the lifetime 
for decaying into two particles.

One can obtain the one-loop expression for $\Delta M^2$ starting from the zero and one loop
expressions for the four graviton amplitudes. This was derived in \cite{IK} and 
we will briefly review this computation in the next Section 2 . We get an 
integral expression in terms of theta functions.
[In \cite{IK} a computation of the lifetime was also attempted; however
the algorithm employed in this first investigation was not sufficiently accurate
and we find now a different and more precise result].    

One finds that $\Delta M^2$ is formally expressed 
 as a divergent integral of a positive quantity:
its imaginary part can be computed by analytic continuation in $M^2$, starting 
from $M^2<0$
where the integral is convergent. By a systematic expansion of the  integrand 
the calculation then
 reduces to a sum of integrals of the form
\beq
I(\alpha ,\omega )=\int ds\ s^{-\alpha}\ e^{sM^2\omega}\ ,
\label{integral}
\eeq
where each term has a multiplicity depending on $\alpha$ and $\omega$. The imaginary part  is
computed by a standard formula:
\beq
\Im \ I(\alpha ,\omega )={\pi (M^2\omega )^{\alpha -1}\ov\Gamma (\alpha )}\ .
\label{imaginary}
\eeq
One finds that this is precisely the expression for the one loop mass-shift in $\phi^3$ field theory,
and that $\omega$ is determined by the masses $M_1$ and $M_2$ of the decay products,
whereas $\alpha$ is related to their orbital angular momentum, which is necessary to 
compensate the mismatch of $J$,$J_1$ and $J_2$. This is reviewed in Section
3.

The difficult task is to compute the multiplicity of $I(\alpha ,\omega )$, due to the 
large multiplicity
of the decay products. We have found a very efficient algorithm for doing that, generalizing the
well known saddle-point technique for computing the multiplicity (i.e. entropy) 
of the string states.
The generalities are presented in Section 4. 
The computation of the imaginary part is presented in Section 5.

In Section 6  we make a numerical analysis of our formulae. This analysis is very fast and accurate.
We have made in particular a nontrivial check: $\Delta M^2$ can be expressed in two 
formally equivalent ways in terms of two different theta functions, which lead to a very 
different rearrangement of terms in the expansion and to different 
saddle point analyses. We find nevertheless identical results, summarized
in a three dimensional plot, see figures 1 and 4.

We  find that ${\rm Im}(\Delta M^2)$ can be written as a sum over the contributions 
of different channels, according to the masses $M_1,M_2$ of the
particles of the decay product.
It turns out that
most of these processes are exponentially suppressed, except
for a line in the space $M_1,M_2$, where the decay rate has a power-like dependence on the coupling. 
This is a surprise, since
we do not see any special reason why, given two masses in a three-state vertex,
there should be a selection rule for the third.
It is likely that that feature results from the interplay among the decay products
entropy and centrifugal barrier and phase space effects
\footnote{We thank Daniele Amati for a discussion on that point.} .


The other surprise is that we find that the  
lifetime of the excited state of maximal angular momentum 
for decaying into two particles grows with the first power of its mass $M$.
This is computed in Section 7.

A final remark:
at tree level the whole closed string decay can be described as a sequence 
of successive two-body decays. 
If the decay products are narrow resonances
 --~particles with a lifetime much larger than the  inverse of their 
mass~--
then the lifetime of the first decay is the lifetime of the massive state 
$tout \   court$, 
like for instance the lifetime of a $K$-meson is set by its decay  into  
$\pi$-mesons, 
despite the $\pi$-meson being also unstable.

\section{The Mass Shift}
Here we review the derivation of the one-loop expression for $\Delta M^2$ 
given in \cite{IK} 
by factorizing the one-loop four graviton amplitude 
obtained  by \cite{Schwarz:1982jn} in closed superstring theory
(type IIA and type IIB give the same result):
\beq
A_{1}=R^4\int{d^2\tau\ov (\Im\tau )^2}\int\prod_{i=1}^3 
{d^2z_i\ov \Im\tau} \ e^{-2\sum_{i<j}^3 k_i . k_j \chi (z_{ij})}
\label{sch}
\eeq
where
\beq
\chi(z)=\log \bigg| {\theta_1 (z|\tau )\ov\theta^{'}_1 (0|\tau )}\bigg|^2
-2\pi{(\Im z)^2\ov\Im\tau}
\eeq
and $R^4$ is a kinematical factor containing the graviton polarizations.
We also set $\al ' M^2=4 N$, and choose units where $\al ' =4$, so that
in what follows we can set $M^2=N$.

The amplitude $A_1$ has a double pole for $S\to N$ ($S=2k_1.k_2$ ) due to the propagator of a massive
string state with $M^2=N$, produced by the collision of the two incoming gravitons, and another
similar propagator coupled to the two outgoing gravitons. The residue of the double pole
is proportional to $\Delta M^2$, the (1-loop) mass shift of the massive state, $M^2\to M^2+\Delta M^2$. One has
$$
A_1\to G_{in}{1\ov S-N}\Delta M^2{1\ov S-N}G_{out}\ .
$$
To get  $\Delta M^2$,
 one divides the double pole residue of $A_1$ by the single pole residue of $A_0$,
the tree level four graviton amplitude \cite{Schwarz:1982jn}:
$$
A_0\to G_{in}{1\ov S-N}G_{out}\ .
$$
The poles of $A_1$ occur as singularities of the integrand in (\ref{sch}) for  $z_{12},z_{34}\to 0$.
The integrand behaves as $\sim |z_{12}|^{2S}|z_{34}|^{2S}\cdot F(z_{12},z_{34},z)$, with 
$z={1\over 2}(z_1+z_2)$. One has to look at the terms in the expansion of $F$ that behave as 
$|z_{12}|^{2N-2}|z_{34}|^{2N-2}$. 
Further, in order to  select the state of maximal angular momentum
$J=2N+2$, one looks for the maximal power of $\cos\theta$
--~the angle between the space-momenta of the ingoing and
outgoing gravitons in the c.m. frame~-- both in the residues 
of $A_1$ and $A_0$. 
In this way one obtains
\be
\Delta M^2  
        & =  &  c    \int { ds d\tau_1 \over s^2}\int {d^2 z \over s}
	 e^{-{4N\pi ^2 y^2\over s}}
            \bigg| {\pi \theta_1(z|\tau )\ov \theta_1'(0|\tau )}\bigg|^{4N} s^{-2N}
\nonumber \\
 & & \times\  \sum_{l=0}^{N-1}{N!^2 \over l!^2(N-l-1)!^2} 
           \left| {s\over\pi ^2 }\p^2_z \log\theta_1(z|\tau )+1\right|^{2l}
\ ,
\label{prima}                                  
\ee
where $s=\pi\Im (\tau )$ and $c$ is a numerical constant, independent of $N$.
In \cite{IK} the normalization was checked by evaluating the contribution to
 $\Im\Delta M^2$ from the decay rate of the excited
state into two massless states, and finding agreement with the explicit computation
of the decay into two gravitons.

\section{Field Theory}

Let us first consider the case of $\phi^3$ field theory.
Consider the one-loop correction to the propagator of a particle of mass
$M$, due to the contribution of particles of masses $M_1$, $M_2$ running
in the loop.
With a convenient parametrization,  the Feynman diagram has the following form
\beq
\Delta M^2
=\sum_{M_1,M_2,l_0} P
\int_0^\infty  ds\ s^{-\beta (l_0)} \ \int_0^1 d\eta\  e^{4Ws }\ ,\ \ \ 
\label{funo}
\eeq
with
$$
 W(\eta )=M^2 \eta (1-\eta)-\eta M_1^2-(1-\eta)M_2^2 \ .
$$
Here $\beta (l_0) =D/2-1+l_0$ ($D=$ spacetime dimension), and 
the polynomial $P=P(M,M_1,M_2,l_0)$
takes into account in particular the multiplicity of the decay products,   
$l_0$ being related to their orbital angular momentum needed for matching the
angular momentum of
the decaying state. 

The IR region is $t=\infty $.
This integral is convergent below the threshold for particle production.
Above the threshold, the integral is defined as usual by analytic
continuation, which gives rise to an imaginary part.

The threshold appears when $W(\eta )$ changes sign and 
becomes positive. The maximum of $W(\eta )$ is at $\eta =\eta_0$, with
$$
\eta _{0}={M^2-M_1^2+M_2^2\over 2M^2}\ ,\ \ \ 
$$
where
$$
W(\eta_{0})=
{1\over 4M^2} \big( M^2-(M_1+M_2)^2 \big) \big(
M^2-(M_1-M_2)^2\big)  = \vec p^{\ 2}\ .
$$
Hence $W(\eta_{0})>0$ for $M^2>(M_1+M_2)^2$. 

For future use we define
\beq
\omega\equiv{4W(\eta_{0})\ov M^2}=1-2(\sigma_1 +\sigma_2 )+
(\sigma_1 -\sigma_2 )^2\ ,\ \ \ \ 
 \sigma_{1,2}={M^2_{1,2} \over M^2}\ .
\label{waza}
\eeq

Since $4W=M^2\omega - 4M^2(\eta-\eta_0)^2 $,
for large $M^2$, we can 
evaluate the integral over $\eta$ by performing a Gaussian integration 
around the maximum at $\eta=\eta_0 $.  Ignoring constant factors, we get
\beq
\int_0^1 d\eta\  e^{4Ws }\sim {1\ov\sqrt{M^2s}}\ e^{s\omega M^2}\ .
\label{etas}
\eeq
We then evaluate the imaginary part of $\Delta M^2$ by analytic continuation from
$M^2\omega <0$ to $M^2\omega >0$ by means of the formula 
seen in eqs. (\ref{integral}) and (\ref{imaginary}).
The final  result  is symmetric in $M_1\leftrightarrow M_2$.

In the case of string theory, we will obtain $\Delta M^2 $
expressed in two different (but equivalent) ways, as discussed in section 4.
For comparison with the string theory expression studied in detail in
appendix A, we set $M^2=N$ and
 make the change of variable
$\eta ={\pi\over s} y$ to get
\beq
 \Delta M^2
=\sum_{M_1,M_2,l_0} P
\int_0^\infty  ds\ s^{-\beta (l_0)-1} e^{-4sM_2^2}\
 \int_0^ {s\over \pi } dy\  e^{{-{4N\pi^2y^2\over s} +
4\pi y(N-M_1^2+M_2^2)}  }\ .\ \ \ 
\label{express1}
\eeq
In order to compare with the string theory expression studied in section 5,
we make another change of variable :
$\eta={\pi\over s} y + {1\over 2}$, getting the field theory expression 
\be
\Delta M^2
& = &
\sum_{M_1, M_2,l_0}  P\int_0^\infty  ds\ s^{-\beta (l_0)-1} e^{[N-2(M_1^2+M_2^2)]s}\
\nonumber\\
& \times &
\int_{-{s\over 2\pi} }^{s\over 2\pi } 
dy\ e^{-{4N\pi^2y^2\over s} - 4\pi y(M_1^2-M_2^2)}\ .  
\label{express2}
\ee

\section{General Method}
Consider the formula for the one-loop string diagram eq.~(\ref{prima}),
which is expressed in terms of $\theta_1 (z|\tau )$.

We note that the integrand can be expanded in a sum of terms
of the form
\beq
T(m_1,m_2)=
s^{-2N+(m_1+m_2)-3}e^{-{4N\pi ^2 y^2\over s}}\cdot Q_{m_1}\cdot\bar Q_{m_2}
\label{T}
\eeq
where
\beq
Q_{m}=
({\pi \theta_1(z|\tau )\ov \theta_1'(0|\tau )})^{2N}({1\ov\pi^2}\p^2_z \log\theta_1(z|\tau ))^m
\label{Q}
\eeq
can be further expanded in powers of $q^2=e^{i2\pi\tau}$ and 
in a Laurent series in $p=e^{2i\pi z}$, which is symmetric under $p\to p^{-1}$.
After the integration over $\Re (\tau )$ and $\Re (z )$ (ensuring $L_0=\bar L_0$ on the states),
we get a sum of terms like
\beq
s^{-2N+(m_1+m_2)-3}e^{-{4N\pi ^2 y^2\over s}}e^{-4 \tilde ks}e^{4\tilde j\pi y}\ ,
\label{bebe}
\eeq
with $y=\Im z$, $s=\pi \Im (\tau )$.

By comparing with eq.~(\ref{express1}) 
we see that $2N-(m_1+m_2)-2$ corresponds to $l_0$, $\tilde k$ to $M_2^2$, and $\tilde j$ to 
$M^2-M^2_1+M^2_2$ . 
We can thus determine $P(M,M_1,M_2,l_0)$.

We will actually compute $\Im\Delta M^2$ at fixed $M_{1,2}$ summing over $l_0$,
that is summing over every possible angular momentum and multiplicity
of the decay products.

It is also useful to make a shift $z\to z+\tau /2$ and  
to re-express the string one-loop diagram in the form
\be
\Delta M^2  
        & =  &  c    \int { ds d\tau_1 \over s^2}\int {d^2 z \over s}
	 e^{-{4N\pi ^2 y^2\over s}}
   e^{Ns} \bigg| {\pi \theta_4(z|\tau )\ov e^{-i\pi\tau /4}\theta_1'(0|\tau )}\bigg|^{4N} s^{-2N}
\nonumber \\
 & & \times\  \sum_{l=0}^{N-1}{N!^2 \over l!^2(N-l-1)!^2} 
           \left| {s\over\pi ^2 }\p^2_z \log\theta_4(z|\tau )+1\right|^{2l}\ .
\label{primab}                                  
\ee

Now we can expand the integrand of eq.(\ref{primab}) in a sum of terms of
the form 
\beq
T'(m_1,m_2)=4^{-2N}
s^{-2N+(m_1+m_2)-3}e^{Ns}e^{-{4N\pi ^2 y^2\over s}}\cdot Q'_{m_1}\cdot\bar Q'_{m_2}\ ,
\label{T'}
\eeq
where now:
\beq
Q'_{m}=
({2\pi \theta_4(z|\tau )\ov q^{-1/4}\theta_1'(0|\tau )})^{2N}
\bigg( {1\ov \pi^2}\p^2_z \log\theta_4(z|\tau )\bigg)^{m}\ .
\label{Q'}
\eeq
Note that the new $Q'_{m}$ has integer power expansions in $q$ and $p$.
By looking at the terms  
$|q|^{2k}=e^{-2ks}$ and $|p|^{2j}=e^{-4j\pi y}$ (after integrating over $\Re (\tau )$ and $\Re (z)$)
and comparing with eq.~(\ref{express2}) we now identify $k=M_1^2+M^2_2$ and $j=M_1^2-M^2_2$.

The two forms of $\Delta M^2$, eq.~(\ref{prima}) and eq.~(\ref{primab}) 
are 
equivalent. However, 
the final expressions that we will obtain using  as starting points 
eq.~(\ref{prima}) and eq.~(\ref{primab}) and computing
integrals by saddle-point approximation will involve very different
expansions.
Therefore it will be a nontrivial
check to verify that indeed one gets the same result.

\section{Calculation of the imaginary part of $M^2$}

Here we evaluate $\Im\Delta M^2$ using the string loop expression
(\ref{primab}), written in terms of $\theta_4$. 

It should be remembered that the result 
for $\Delta M^2$   does not change if we 
replace $\theta_4 $ by $\theta_{1}$, $\theta_2 $ or $\theta _3$, 
since they differ by a shift of $z$ and a factor that compensate the change from
$ e^{-{4N\pi ^2 y^2\over s}}$. 
In appendix A we repeat the analysis with $\Delta M^2$  expressed
 in terms of $\theta_1$,
eq.~(\ref{prima}).

We expand the binomial
$({s\over\pi ^2 }\p^2_z \log\theta_4(z|\tau )+1)^l$ and using the formula
$$
\sum_{l={\rm max}(m_1,m_2)}^{N-1} {1\over (l-m_1)!(l-m_2)!(N-l-1)!^2}={(2N-m_1-m_2-2)!\over
(N-m_1-1)!^2 (N-m_2-1)!^2}
$$
we get (with $\tau =\tau_1 +i \ s/ \pi $
and $z=x+i y$)
\be
\Delta M^2 &=&c \int ds d\tau_1\int dydx 
\nonumber\\
&\times & \sum_{m_1,m_2}T'(m_1,m_2)
{(2N-m_1-m_2-2)!N!^2\over m_1!(N-m_1-1)!^2 m_2!(N-m_2-1)!^2}
\ee
with $T'(m_1,m_2)$  expressed in terms of $Q'_{m_{1,2}}$  as in eq.(\ref{T'}). 
Now we expand 
\beq
 Q'_m= \sum_{k,j} c_{kj}(N,m)\ q^k p^j\ ,\ 
 \eeq
 with
\beq\ \ \ 
 c_{kj}(N,m)=-{1\over (2\pi)^2}\oint {dq\over q} \oint {dp\over p}
 q^{-k} p^{-j} Q'_m\ ,
\eeq
  and similarly for the complex conjugate.
{} From the explicit expressions for the $\theta_4 $ 
function given in sect.~6, one can see 
that the sum over $k$ contains only positive integer values of
$k$, whereas the sum over $j$ contains both positive and negative
powers of $j$, with the property $c_{kj}=c_{k(-j)}$.

Let us now consider the integrals over $\tau_1$ and $x$.
Since the imaginary part of $\Delta M^2$ comes from the divergence of the integral at $s\to \infty $,
we can replace the integral over the fundamental domain by an integral over the full strip.
In addition, we note that $\theta_4\to \theta_3 $ by a shift $\tau_1 \to \tau_1+1$. Given that the original integral gives the same result
for $\theta_3 $ and $\theta _4$, we can extend the integration region in $\tau_1$ to the interval $(-1,1)$.
Then
\be
{1\over 2}\int_{-1}^1 d\tau_1 \int _0^1 dx\ Q'_{m_1}\bar Q'_{m_2}
&=& 
\sum_{k,j} (q\bar q)^k (p \bar p)^j  c_{k j}(N,m_1)\bar  c_{k j}(N,m_2)
 \nonumber\\
 &=& 
 \sum_{k,j} e^{-2 k s-4\pi j y}\  c_{k j}(N,m_1)\bar  c_{k j}(N,m_2)
\nonumber
\ee
The integration over $y$ is performed by saddle point as
in eq.~(\ref{etas}). Then we consider the integral over $s$,
and  use the general formula ({\ref{imaginary}) 
for computing the imaginary part. We obtain
\be
{\rm Im} (\Delta M^2  )
&  \sim &
{ 4^{-2N}\ov \sqrt{N}}  
\sum_{j,k} (N\omega )^{2N+{3\over 2}}
\sum_{m_1,m_2=0}^{N-1}\sum_{m_1,m_2=0}^{N-1}{(2N-m_1-m_2-2)!\ov \Gamma(2N-m_1-m_2+{5\over 2})} 
\nonumber \\
&\times &
{N!^2\ c_{kj}(N,m_1)\ \bar c_{kj} (N,m_2)\ (N\omega)^{-m_1-m_2}
\over m_1!(N-m_1-1)!^2 m_2!(N-m_2-1)!^2} 
\label{brutta}
\ee
Here
$$
\omega (\rho ,\sigma  ) \equiv  1- 2 \s +\rho^2\ ,
$$
and
\beq
\sigma \equiv  {k\over N}\ ,\ \ \ \ \rho \equiv {j\over N} \ .
\label{yty}
\eeq
We remind that by comparing with field theory we see that the integers $k, j$ are related to the
masses $M_1, M_2$ of the decay product (cf.~eq.~(\ref{waza})~):
\beq
k=M_1^2+M_2^2\ ,\ \ \ \ \ j=M_1^2-M_2^2\ .
\label{kajo}
\eeq

Here we consider large values of $N$ with fixed $\s $ and $\rho $. Other cases
will be discussed in appendix B and C.

It will be clear from the calculation below that the main
 contribution in the sum over $m_1,m_2$ comes
 from the region where $2N-m_1-m_2$ is large.
Therefore we can approximate
\beq
{(2N-m_1-m_2 -2)! \ov \Gamma(2N-m_1-m_2 +{5\over 2}) }\sim N^{-7/2}(2-{m_1\over N}-{m_2\over N})^{-7/2}\ .
\eeq
Moreover, in the large $N$ limit, the sum over $m_{1,2}$
 is dominated by a sharp maximum; away from the maximum 
the terms are exponentially suppressed like $e^{-cN}$.
We will see that on the maximum
$r(\rho ,\sigma )\equiv (2-(m_1+m_2)/N)^{-7/2}$ is a finite function of $j/N,k/N$. 

Therefore,  we can write
\beq
{\rm Im} (\Delta M^2  )
 \sim    4^{-2N}N^{2N-1/2}\sum_{j,k} \omega ^{2N+{3\over 2} }\  
r(\rho ,\sigma )\ |L(j,k)|^2\ ,
\label{sumjk}
\eeq
where
\beq
L(j,k)= \oint {dq\over q} \oint {dp\over p}
 q^{-k} p^{-j} \left(
{2\pi \theta_4(z|\tau )\ov q^{-1/4}\theta_1'(0|\tau )}\right) ^{2N} H(q,p;j,k)\ ,
\label{L}
\eeq
and in turn
\beq
H(q,p;j,k)=  
\sum_{m=0}^{N-1} {(N-1)!\ov m!(N-m-1)!^2 } \left( {{1\ov \pi^2}\p^2_z \log\theta_4(z|\tau )\ov N\omega }\right) ^{m}\ .
\label{H}
\eeq

\section{ Numerical evaluation of $ {\rm Im} (\Delta M^2 )$}


We now determine the  functions appearing in (\ref{sumjk}).
We  will make use of the formulas in terms of  $q=e^{i\pi\tau_1-s}$ and $p=e^{i 2\pi z}$: 
\be
{2\pi \theta_4(z|\tau )\over  q^{-1/4}
\theta'_1(0|\tau )} &=&  \prod _{n=1}^\infty
 { (1- p q^{2n-1} ) (1- p^{-1} q^{2n-1} ) 
\over
 (1-q^{2n})^2} \nonumber\\
& \equiv & \exp \big[ {1\over 2} f(q,p) \big]\ .
\ee
By expanding the logarithms in the above definition of $f(q,p)$, 
and interchanging the two infinite sums,
we obtain
\beq
f(q,p)=-2\sum_{n=1}^\infty {q^n (p^n+p^{-n}-2 q^n)\over  n (1-q^{2n}) }\ .
\eeq

Further:
\beq
{1\ov \pi^2}\p^2_z \log \theta_4(z|\tau) =
4 \sum_{n=1}^\infty {q^{2n-1}\big[ (p+p^{-1})(1+q^{4n-2})-4 q^{2n-1}\big]\over
(1-pq^{2n-1})^2(1-p^{-1} q^{2n-1})^2 }\equiv   g(q,p) 
\label{ggg}
\eeq

Define
$$
v \equiv {g(q,p)\over \omega(\rho ,\s ) }\ .\ \ \
$$
Now we use the formula (\ref{H}) for $H=H(q,p,j,k)$ : 
\beq
H =  \sum_{m=0}^ n
{ n! \over m! (n-m)!^2 }  ({v\over n+1})^{m}
 =  ({v\over n+1})^{n} {1\over 2\pi i} \oint 
{dt\over t^{n+1}}   e^{(n+1) t} (t+{1\over v})^{n}  \ .
\eeq
with $n=N-1$. 
{}For large $N$, the integral over $t$ 
can be computed by a saddle point evaluation.
We get
\beq
H(v,N) \cong N^{-N+1/2} \exp\big[N h(v) \big]\ ,
\label{hha}
\eeq
where
\beq
h(v)=  \log v+\log{\sqrt{1+4v}+1\over \sqrt{1+4v}-1} +
{1\over 2v}(\sqrt{1+4v}-1)\ .
\label{wera}
\eeq
We have checked  that this formula provides  a
very accurate representation
for the sum in (\ref{H}) 
already for $N$ larger than 10. 

Thus we finally get
\beq
L(j,k)=N^{-N+1/2} \oint {dq\over q} \oint {dp\over p}
 q^{-k} p^{-j} e^{N \big[ f(q,p)+ h(v(q,p))\big]}\ .
\label{saa}
\eeq

The remaining integrals over $q $ and $p$ can also be computed by a 
saddle-point evaluation.
Since the functions appearing in the integrand are complicated,
it is more convenient to perform this calculation by a numerical evaluation.

The saddle-point evaluation of the integrals over $q, p$ is done
numerically by first finding the extremum of the exponent in eq.~(\ref{saa}).
This determines
$$
q_0=q_0(\rho ,\s)\ ,\ \ \ \ \ p_0=p_0(\rho ,\s)\ .
$$
We find numerically that the saddle point is obtained for $q_0$ and $p_0$ real and positive.
By performing the Gaussian integration around the saddle point we get
\beq
|L(j,k)|^2\sim N^{-2N+1-2}e^{2NS_L(\rho ,\sigma )}\ ,
\label{saddle}
\eeq
where 
\beq
S_L(\rho ,\sigma )=-\sigma\log q_0-\rho\log p_0+ \Re [f(q_0,p_0)+ h(v(q_0,p_0))]\ .
\label{esuno}
\eeq
It is seen that the saddle is a minimum of $S_L$ for $q_0,p_0$ real, and that on it
$f$ and $h$ are real.

Finally, from eq.(\ref{sumjk}) we obtain
\beq
{\rm Im} (\Delta M^2  )
 \sim    N^{-3/2}\sum_{j,k} \omega^{3/2}(\rho,\s )\ r(\rho,\s ) \ e^{2NS_0}
\label{final}
\eeq
where 
\beq
S_0(\rho ,\sigma )=S_L(\rho ,\sigma )+\log \omega (\rho ,\sigma ) -\log 4 \ .
\label{esdos}
\eeq

Also, we mention that
the result for (\ref{wera}) can also be obtained by evaluating the sum over $m$ in eq.(\ref{H})
by looking at the maximum, found for $m_0=N(1-{1\over 2v}(\sqrt{1+4v}-1))$, and expanding around it
(this is done in the Appendix C).

In particular, on the maximum   $(2-{m_1\over N}-{m_2\over N})={1\over v}(\sqrt{1+4v}-1)$, since the same
maximum holds for $m_{1,2}$. 
Therefore we find
\beq
r(\rho ,\sigma )= \left( 
{1\over v(q_0,p_0)}(\sqrt{1+4v(q_0,p_0)}-1) \right) ^{-7/2}\ .
\label{r}
\eeq
 

Our final formula for the rate of the decay channel to particles of masses $M_1,M_2$ 
is thus given by
\beq
\Gamma(M_1,M_2)= {1\over 2\sqrt{N} }\ \Im\Delta M^2\bigg|_{M_1,M_2}  =
  \omega^{3/2}\ r\ 
{1\over  N^{2} }\ e^{ 2N S_0}\ ,
\label{ampp}
\eeq
where $S_0$ is a function of the ratios ${M_1\over M},\ {M_2\over M}$, with $M=\sqrt{N}$.

Note that $\Gamma(M_1,M_2)$ represents the contributions of all decay 
channels involving  particles
with the same masses $M_1,M_2$ (since the multiplicity grows exponentially with the mass,
for large $M_1,M_2$, there is an exponentially large number of particles
contributing to $\Gamma (M_1,M_2)$).

Figure 1 is a numerical plot of $S_0$ in function of $M_1$ and $M_2$.
We see that $S_0$ is negative definite, except on some curve (see fig.~3)
where
it identically vanishes.
Thus the first observation is that
the  rate for the  decay channel to particles of {\it generic} 
masses $M_1,\ M_2$
is  exponentially suppressed at large $N$.

In the Appendix A, figure 4 shows the numerical plot of $\tilde S_0$, 
obtained by starting 
with the expression (\ref{prima}) in terms of $\theta_1$. The two figures are identical.

\begin{figure}[hbt]
\label{fig1} \vskip -0.5cm \hskip -1cm
\centerline{\epsfig{figure=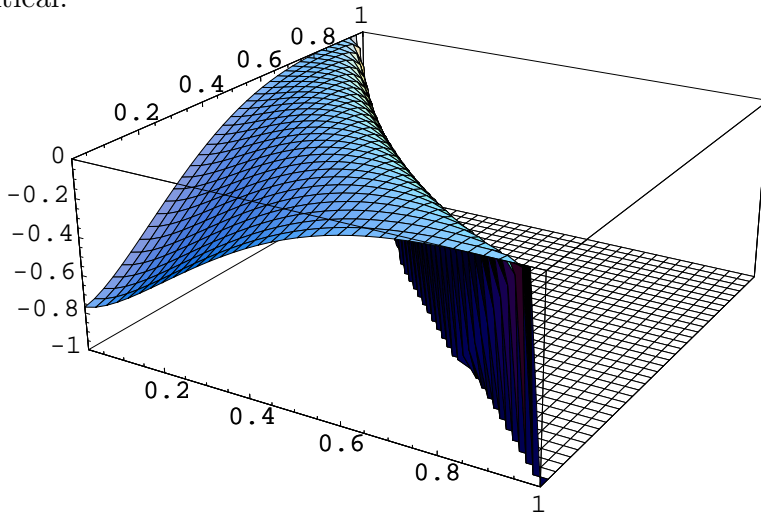,height=7truecm}}
\caption{\footnotesize 
The exponent $S_0$ of the decay rate $\Gamma (M_1,M_2)$ 
in terms of the masses $M_1^2/\sqrt{N}$, $M_2/\sqrt{N}$,
computed using the formula (\ref{esdos}).
The kinematically allowed region is
inside the triangle defined by $M_1,M_2>0$, ${M_1\over M}+{M_2\over M}<1$.
The maximum of $S_0$ is $S_0=0$ and it is located on a curve shown in figure 3.}
\end{figure}

Figure 2 is a plot of $S_0(M_1)$ for  given $M_2$, i.e.  slices of
figure 1 at constant $M_2$.
One can see that the maximum exactly passes by $S_0=0$. This happens for
any $M_2$. We have checked that the factor $\omega^{3/2}(\rho,\s )\ r(\rho,\s )$
in eqs.(\ref{final}) and (\ref{ampp}) is finite inside the allowed triangle,
except on the boundary $M_1+M_2=M$ where $\Gamma$ is anyhow suppressed.

Modulo the exponentially suppressed processes,
a massive particle will decay  through the special channel
where $S_0$ vanishes.
This defines a curve $M_2=F(M_1)$ in the space $M_1,M_2$, which is shown in 
figure 3.
Such dominant channels exhibit a power-like behavior
\beq
\Gamma (M_1,M_2)\sim N^{-2}\ .
\label{gamma}
\eeq
It is  remarkable that for large $N$ the dynamics ``excludes" 
 decays into kinematically allowed channels.
{} In other words, we find that if the massive particle decays into
a particle of mass $M_1$, the mass of the other particle $M_2$
is uniquely determined, modulo exponentially suppressed processes.

\begin{figure}[hbt]
\label{fig2} \vskip -0.5cm \hskip -1cm
\centerline{\epsfig{figure=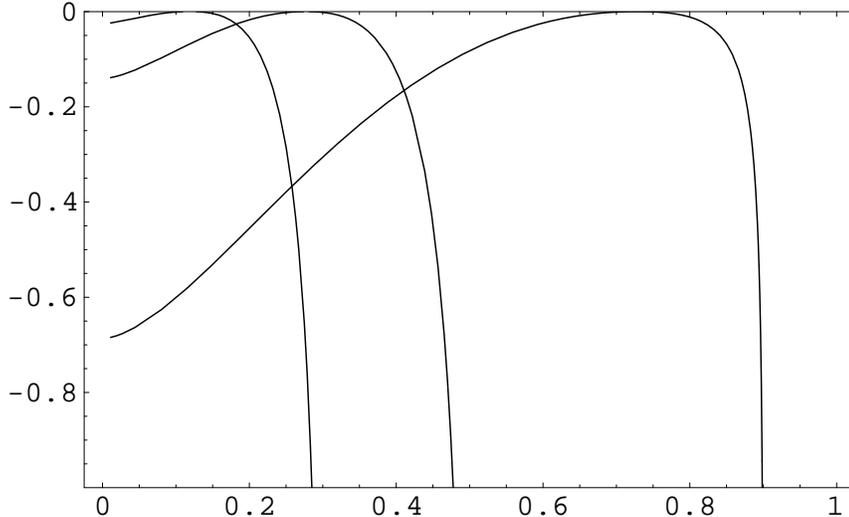,height=7truecm}}
\caption{\footnotesize 
Sections 
of figure 1 at constant $M_2$. The three curves
displayed corespond to ${M_2\over M}=0.1, 0.3, 0.7$
 (with maxima located
at ${M_1\over M} \sim 0.74,\ 0.28,\ 0.12$ respectively).
One can see that the
maximum always passes by $S_0=0$ for every value of $M_2$.}
\end{figure}

The curve $M_2=F(M_1)$ is well approximated by
the curve
\beq
({M_1\over M})^{a }+({M_2\over M})^{a }=1\ ,\ \ \ \ \  a \cong 0.73\ ,
\label{booky}
\eeq
also shown in fig.~3. Although this not the true 
analytical formula connecting  $M_1, M_2$ 
(which is extremely complicated), eq.~(\ref{booky}) is useful 
as a book-keeping of the approximate relation between $M_1$ and $M_2$.

\begin{figure}[hbt]
\label{fig3} 
\vskip -0.5cm \hskip -1cm
\centerline{\epsfig{figure=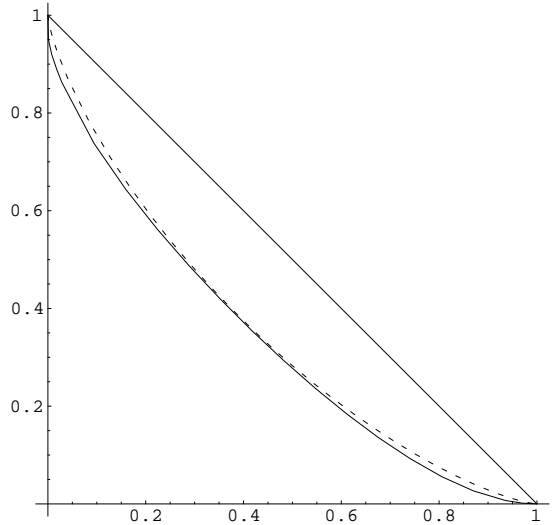,height=7truecm}}
\caption{\footnotesize 
The curve defined by $S_0(M_1,M_2)=0$ (solid line), describing the dominant channels.
For these special values of $M_1,M_2$ the rate is not
exponentially suppressed.
The dashed line is the curve $(M_1/M)^{a}+(M_2/M)^{a}=1$,\ 
$a = 0.73$.}
\end{figure}

In the Appendices B and C we have considered the cases when one 
of the masses $M_{1}$ or $M_{2}$, or both,
are small with respect to $N$. We find that the only case which is not exponentially suppressed
is when $ M_2=0$ and $M_1^2=N-j$ with $j$ finite (or $viceversa$). In this case
$$
\Gamma \sim N^{-5/2} \ .
$$
Thus the dominant channel is for both $M_{1,2}^2$ of order $N$ along the curve $M_2=F(M_1)$,
where the decay width is given by eq.(\ref{gamma}). 

\section{Lifetime of a state with maximum angular momentum}

Writing the formula (\ref{final}) for  $\Im (\Delta M^2 )$ in terms of  $\s_{1,2} =M_{1,2}^2/N $,
 we obtain
\beq
\Im (\Delta M^2 )= N^2\int_{ {\cal D} } d\s_1 d\s_2  
{1\over N^{3/2}}\ e^{2N S_0}\ ,
\label{pas}
\eeq
where the domain of integration is the region $\s_1 ,\s_2>0$ 
and $\sqrt{\s_1 }+\sqrt{\s_2}<1$.

We have seen that  $S_0$ is in general negative except 
on the curve where $S_0$ vanishes. Thus only a small  neighborhood of this curve
contributes to the integral
(\ref{pas}). Let $ l$ be a parameter along the line, $0< l<1$,
and let $ n$ a parameter for the orthogonal direction, where $ n=0$
means a point on the curve. It is convenient to use $ n,\  l$ as integration variables. The integration over $ l$ is trivial, since
$S_0$ takes the same value (i.e. equal to zero) for all $ l$.
In the vicinity of the line, we can expand $S_0$
in powers of $ n$, and get a Gaussian integral over $n$
of the form 
\beq
\Im (\Delta M^2 )= N^2\int d n\ {1\over N^{3/2}}\ e^{-c N n^2}\ ,
\label{passs}
\eeq
where $c$ is a number of order 1. The Gaussian integral gives
an extra factor $N^{-1/2}$, so we get
 (see eq.~(\ref{aaq})~)
\beq
\T ^{-1}= {\Im (\Delta M^2 )\over 2\sqrt{N}}={\rm const.}\ {1\over\sqrt{N}}
\eeq
or
\beq
\T= {\rm const}\ \al' M\ .
\eeq
where we have restored $\al' $.
Thus the lifetime of a state with maximum angular momentum
in closed superstring theory is proportional to the mass.

\bigskip\bigskip

\noindent {\bf Acknowledgements}:
We would like to thank D. Amati for a useful discussion.
R.I. acknowledges partial support by the European network
HPRN-CT-2000-00131.
J.R. wishes to thank the Abdus Salam 
International Centre for Theoretical Physics for hospitality while
part of this work was carried out.

\setcounter{section}{0}
\setcounter{subsection}{0}


\appendix{Alternative calculation of $\bf {\rm Im}(\Delta M^2)$}

Here we provide an alternative calculation of  ${\rm Im}(\Delta M^2)$ 
by using as starting point the formula (\ref{prima}) 
in terms of $\theta_1$.

The computation is quite similar to the one done in Sects.~5 and 6, 
this time using the expansion of 
eqs.(\ref{T}) and (\ref{Q}).
We also make use of the formulas
\be
{\pi \theta_1(z|\tau )\over \theta'_1(0|\tau )} &=& \sin \pi z
 \prod _{n=1}^\infty
 { (1- p q^{2n} ) (1- p^{-1} q^{2n} ) 
\over
 (1-q^{2n})^2} \nonumber\\
& \equiv &  \exp \big[ {1\over 2} \tilde f(q,p) \big]
\label{auno}
\ee
\be
{1\ov\pi^2}\p^2_z \log \theta_1(z|\tau) & = &
{4\over p+p^{-1}-2} + 
4 \sum_{n=1}^\infty {q^{2n}
 [(p+p^{-1})(1+ q^{4n})- 4 q^{2n}]\over
(1-pq^{2 n})^2(1-p^{-1} q^{2 n})^2} 
 \nonumber\\
& \equiv  & \tilde g(q,p)\ .
\label{aggz}
\ee
By expanding the logarithms in the above definition of $\tilde f(q,p)$,
and interchanging the two infinite sums,
we now obtain
\beq
\tilde f(q,p)=\log [{1\over 4} (p+p^{-1}-2)]-
2\sum_{n=1}^\infty {q^{2 n} (p^n+p^{-n}-2 )\over  n (1-q^{2n}) }\ .
\eeq
Note that $\tilde f$ can be defined modulo $i\pi$, that is modulo the sign inside the logarithm,
since in eq.(\ref{primab}) only $\exp{(2\Re \tilde f)}$ appears. 

Both $\tilde f$ and $\tilde g$ are even functions of $q$.
We get the same formulas (\ref{L}) , (\ref{H}), (\ref{hha}) and (\ref{wera}), 
with $f\to \tilde f$ and $g\to \tilde g$
and thus $v\to\tilde v$.
Moreover, as explained in Sect.~4, the relation with the masses $M_{1,2}$ 
of the exponents $\tilde j,\tilde k$ of the expansions $q^{2\tilde k}$, $p^{\tilde j}$  
is now different.  In terms of the exponents $j,k$ of Sect.~5
we have 
$$
\tilde k=(k-j)/2 \ , \ \ \ \ \ 
\tilde j=M^2-j\ .
$$
In conclusion we obtain
\beq
{\rm Im} (\Delta M^2  )
 \sim    N^{-3/2}\sum_{j,k} \omega ^{3/2}(\rho,\s )\ \tilde r(\rho ,\sigma )\ 
e^{2N\tilde S_0(\rho ,\sigma )}\ ,
\label{tildez}
\eeq
where 
\be
\tilde S_0(\rho ,\sigma )
&=& -(\sigma -\rho )\log \tilde q_0-(1-\rho )\log \tilde p_0
\nonumber\\
&&+\Re [\tilde f(\tilde q_0,\tilde p_0)
+ h(\tilde v(\tilde q_0,\tilde p_0))]+\log \omega(\rho,\s ) \ ,
\label{yuo}
\ee
$$
\sigma\equiv {M_1^2+M_2^2\over N}\ ,\ \ \ \ \ \rho\equiv {M_1^2-M_2^2\over N}\ ,
$$
and the saddle point values $\tilde q_0,\tilde p_0$ correspond to the stationary point of $\tilde S_0$.

\begin{figure}[hbt]
\label{fig4} \vskip -0.5cm \hskip -1cm
\centerline{\epsfig{figure=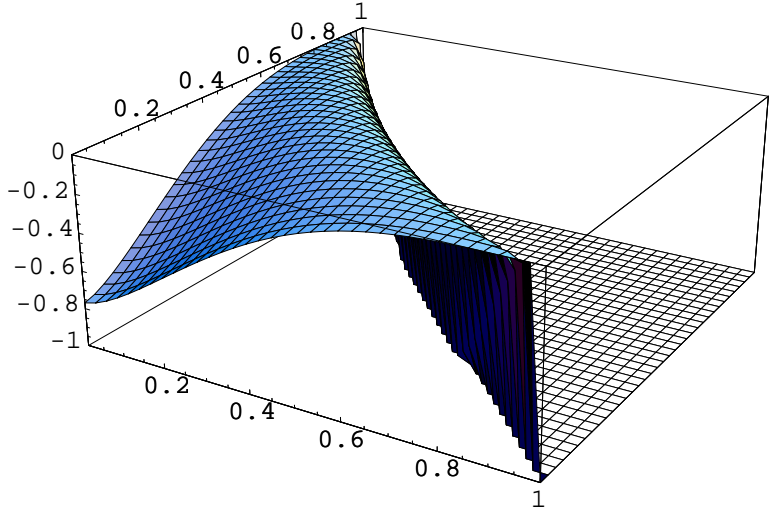,height=7truecm}}
\caption{\footnotesize 
The exponent $\tilde S_0$ of the decay rate $\Gamma (M_1,M_2)$ 
in terms of the masses $M_1/M$, $M_2/M$,
computed using the formula (\ref{yuo}).
The plot should be compared to figure 1, computed by following a
different way.}
\end{figure}

The functions that enter into the final expression  (\ref{yuo})
are individually
 very different from the ones appearing in the formula for
$S_0$ in Sect.~6.
The result is however the same, to a surprising degree of accuracy, 
see fig.~4. 




\appendix{Decay into two massless particles}


The decay rate of the massive particle into two massless particles
(e.g. gravitons) can also be obtained
from the general formula (\ref{primab}).
We need to consider the special channel with $M_1=M_2=0$, i.e.
the powers $q^k p^j$ in the expansion with $k=j=0$.
This is obtained by formally setting $g(q,p)\to 0$ (since it starts with $q^1$)
and ${2\pi \theta_4(z|\tau )\over q^{-1/4}\theta'_1(0|\tau )} \to 1 $.
{}For this $k=j=0$ term, we thus get 
\beq
\Delta M^2\bigg|_{\rm massless}  
         =    c    \int { ds d\tau_1 \over s^2}\int {d^2 z \over s}\ 
	 e^{-{4N\pi ^2 y^2\over s}}
        e^{N s} ({1\over 4 s})^{2N}\ {N^2 (2N-2)!\over (N-1)!^2}\ .
\label{zzz}                                  
\eeq
Computing the integrals over $x, y$ and $\tau_1$, 
we are left with an integral over $s$ whose imaginary part
gives
$$
{\rm Im} \int ds \ s^{-{5\over 2} -2N} e^{Ns} \cong  {1\over\sqrt{N} } e^{2N} 2^{-2N}
$$
Hence
\beq
{\rm Im}\Delta M^2\bigg|_{\rm massless} 
\sim  \sqrt{N}
\ e^{-2 N(\log 4 -1)} \ .
\label{zzzu}
\eeq
This is exponentially decreasing for large $N$.

Note that this decay corresponds to the corner $M_1=M_2=0$ in figure~1,
which is also exponentially decreasing. Indeed, the numerical value
at $M_1=M_2=0$ of the plot is $\cong -0.7726$, which agrees with
$ -2(\log 4-1)$. 

In \cite{IK} this result has been compared with the explicit direct 
computation of the decay into two gravitons, finding complete agreement.

\appendix{Decay into a massless and a massive particle, and remaining cases.}

\def\jj{j }

Another interesting special case is when the massive particle decays into a massless particle
(e.g. graviton) and a massive particle.
So, consider the decay $M \rightarrow M_1+ M_2$ with $M_2=0$ and 
$M_1^2=N- \jj $ (we remind that $M^2=N$).
We consider both the case when $\jj $ is a finite integer and
the case $\jj =c\ N$ where $c$ can be a constant less than one
or $c\sim N^{\lambda -1}$ with $\lambda <1$.
The case $M_2=0$ is most suitably treated by the formula  (\ref{prima})
expressed in terms
of $\theta_1 (z|\tau )$, setting $q=0$ to isolate the term corresponding
to $M_2=0$ in the expression (\ref{bebe}).
Thus (see (\ref{auno}), (\ref{aggz})~)
$$
{\pi\theta_1 (z|\tau ) \over \theta_1^{'} (0|\tau )} \to \sin(\pi z)\ ,
\ \ \ {1\ov \pi^2}\p^2_z \log\theta_1 (z|\tau )\to -{1\ov \sin^2(\pi z)}\ .
$$ 
Therefore now 
\beq
u(p)\equiv g(q=0,p)={4\ov (p^{1/2}-p^{-1/2})^2 }\ ,
\eeq
\beq
e^{f(q=0,p)}=\left( {\pi\theta_1 (z|\tau )\over \theta_1^{'} (0|\tau ) }
\right)^2=-u(p)^{-1}\ .
\eeq
Now $\omega =({j\ov N})^2$ thus $\omega N= {\jj }^2/N$ and
${g\ov \omega N}={u(p)N\ov {\jj }^2}$.
By following similar steps as those which led to
 eq.(\ref{brutta}), now we obtain 
\be
\Im \Delta M^2 &\sim & {(\omega N)^{2N+{3\over 2} }\over \sqrt{N}}
\oint {dp\over p} p^{-\jj }(u(p))^{-N}
\oint {d\bar p\over \bar p}{\bar p}^{-\jj }(u(\bar p))^{-N}
\nonumber\\
&\times & \sum_{m_1, m_2=0}^{N-1}{(2N-m_1-m_2-2)!
\ov \Gamma(2N-m_1-m_2+{5\over 2})}
\ Z_{m_1}\ \bar Z_{m_2}\ .
\ee
where
$$
Z_m\equiv Z(u(p),m,\jj ,N)\equiv
{N!\ov m!(N-m-1)!^2}({u(p)N\ov {\jj }^2})^m \ .
$$
We use a saddle point technique to sum over $m_{1,2}$: 
the dominant contribution comes from the maximum of the 
exponential dependence in those variables. 
Using the Stirling formula, we find
\beq
Z_m\sim (N-m-1)\sqrt{N\over m}\ e^{N\log{N}}
e^{I}\ ,
\eeq
\be
I &=& N-m-2N\log{(N-m-1)}
\nonumber\\
&+&
m \left( -\log{m}+2\log{(N-m-1)}+\log{u(p)N\ov {\jj }^2}\right) \ .
\ee
Imposing that the derivative in $m$ of the exponent vanishes,
we get the equation for $m_0$, the maximum locus,
$$
(N-m_0-1)^2= {m_0 {\jj }^2\ov u(p)N}\ ,
$$
which is solved to give
\beq
N-m_0=1-{{\jj }^2\ov 2Nu}+\sqrt{(1-{{\jj }^2\ov 2Nu})^2+{{\jj }^2\ov u}-1}\ \ .
\label{trew}
\eeq

Let us now discuss different cases.
\medskip

{}For $\jj ^2/N\to 0$ , eq.~(\ref{trew})  gives
\beq
m_0\cong N-{ {\jj }\over \sqrt u } -1\ .
\eeq
On the maximum, we get
$$
Z(u(p),m_0,\jj ,N)\sim 
{{\jj }\ov  \sqrt{u} }
({N\ov {\jj }^2})^N\ u^N\ \exp \big[{\jj \ov\sqrt u} \big]\ ,
$$
and 
$$
{(2N-m_1-m_2-2)!\ov \Gamma(2N-m_1-m_2+ {5\over2} )}\sim 
{\jj }^{-7/2}({1\ov \sqrt{u(p)} }+{1\ov \sqrt{ u(\bar p) } })^{-7/2}\ .
$$
The second derivative of the exponent $I$ is
$-1/m_0-2/(N-m_0-1)$, which in this limit is $ 2\sqrt u/{\jj }$.
We are left  with a Gaussian integral in $\delta m_1, \ \delta m_2$ with a spread of order
$\sqrt{\jj }$, which is small compared to the range of $m$.
Hence
\beq
\Im \Delta M^2\sim {(\omega N)^{2N+{3\over 2} }\ov\sqrt{N}}
({N\ov {\jj }^2})^{2N}\ {\jj }^{-7/2}\  
\bigg| {\jj }^{3/2}\oint {dp\over p} p^{-\jj }
\exp{{\jj \ov \sqrt{u}}} \bigg| ^2\ ,
\eeq
where we  have  neglected finite powers of $u$
and kept into account the range in $m$. 
We recall that $1/\sqrt{u}={1\over 2}(p^{1/2}-p^{-1/2})$. 
The remaining integral over $p$
is obtained again by saddle point technique: defining $x=p^{1/2}$, 
we require the derivative in $x$ of the exponent to vanish, that is
$-2x^{-1}+ {1\over 2}(1+x^{-2})=0$. The solution is  $x=x_0=2+\sqrt{3}$, 
discarding the other solution $x=1/x_0$ which would make 
$N-m\sim {\jj }/u(p)<0$. We get
$$
\oint {dp\over p} p^{-\jj }\exp \big[ {\jj \ov 2}(p^{1/2}-p^{-1/2}) \big]
\sim {1\ov \sqrt{\jj }}\exp \big[-c_0\jj\big]\ , 
$$
where
$$
c_0= -2\log{x_0}+{x_0^2-1\over 2x_0 } \cong 0.9\ .
$$
Thus we finally obtain
\beq
\Im \Delta M^2\sim {{\jj }^{3/2}\ov N^2}\exp \big[ -2 c_0 \jj \big]\ .
\eeq
{}For large $j$, this rate is exponentially suppressed.

The calculation applies as well to the case of finite $\jj $.
In this case, the  exponential factor depending on $j$ is a 
finite number of order $O(1)$, so one obtains
$$
\Gamma \sim N^{-5/2} \ . 
$$

\medskip

Now consider the case when $N>\jj \geq N^{1/2}$. Then eq.~(\ref{trew}) gives 
 $N-m_0-1\sim {{\jj }\ov u } -
{\jj ^2 \over 2Nu} $, 
and therefore we find the same result, up to negligible corrections.

\medskip 

Finally, consider the case $\jj =cN$. Then eq.~(\ref{trew}) reduces to
$$
N-m={1\over 2}({Nc^2\ov u}-2)(\sqrt{1+{4u\over c^2}}-1)\ ,
$$
and we have to evaluate
$$
\oint {dp\over p} p^{-cN}\exp\left[ N \bigg( -\log{u}-
2\log{(\sqrt{1+ {4u\over c^2} }-1)c^2\over 2u}
                             + {c^2\over 2u} 
(\sqrt{1+ {4u\over c^2} }-1)\bigg) \right]\ .
$$
We note that this is a particular case of
the expression of the general case (\ref{saa})
--~we recognize $\tilde v(q=0,p)=u(p)/c^2$ and the expression 
$N(\tilde f(q=0,p)+h(\tilde v(q=0,p))$ in the exponent~--
except that in the general case $M_2^2\sim N$ we have an extra Gaussian integral coming from the integration over $q$ (rather than setting $q=0$ 
to isolate the term corresponding to $M_2^2=0$). As a result,
(\ref{saa}) will give
an extra factor $|1/\sqrt{N}|^2$.

\medskip

Summarizing, for $M_2=0$ and $M_1^2=N-\jj $ with $1\leq \jj  \leq cN$,
we get
\beq
\Im \Delta M^2\sim {{\jj }^{3/2}\ov N^2}\exp\big[ -b \jj \big] 
\label{result}
\eeq
where $b>0$ is a numerical coefficient of order 1.

The case when $\jj =N$, i.e. $c=1$, corresponds to the $M_1=M_2=0$
case treated in appendix B. It can also be recovered as follows.
We need to compute  the coefficient of the power $p^N$ 
in the Laurent expansion of
\newpage
$$
\exp \big[ N\big( -\log{u}-2\log { \sqrt{1+4u}-1\over 2u }
                                +{1\over 2u} (\sqrt{1+4u}-1) \big) \big] 
$$
\beq
=({p\ov 4})^N\ \exp\big[ N(1+\sum_{n>0}c_n p^{-n}) \big] 
\eeq
The coefficient of $p^N$ is $({e\ov 4})^N$. 
Now we have to take the modulus square of this coefficient, 
and take into account that the generic case
eq.(\ref{result}) has an extra factor $|1/\sqrt{\jj }|^2$ originating from the Gaussian integral around the saddle point over $p$.
We thus obtain
the correct result for $M_1=M_2=0$:
\beq
\Im \Delta M^2\sim N^{1/2}\cdot ({e\ov 4})^{2N}\ ,
\eeq 
in agreement with appendix B.

Finally, we have also considered the case when $M_2^2=n$ is 
finite or small with respect to $N$,
which is allowed for $j>2\sqrt{nN}-n$. We find that this case is exponentially suppressed.

\end{document}